\documentclass[12pt,letterpaper, fleqn]{article}

\usepackage{euler}
\usepackage{float}
\usepackage{epstopdf}

\usepackage{titlesec}
\usepackage{amsmath}
\usepackage{amsfonts}
\usepackage{fancyhdr}
\usepackage{enumitem}
\usepackage{xpatch}
\usepackage{amssymb}
\usepackage{amssymb}
\usepackage{amsmath}
\usepackage{amscd}
\usepackage{enumerate}
\usepackage{array}
\usepackage{tabularx}
\usepackage{t1enc}
\usepackage[mathscr]{eucal}
\usepackage{graphicx}
\usepackage[utf8]{inputenc}

\begin{document}

\centerline{\large 
  Nonlinearity of the non-Abelian lattice gauge field theory}
  \centerline{\large according to the spectra of Kolmogorov-Sinai  entropy and}
  \centerline{complexity}
  
\vspace{0.5cm}

\centerline{A. F\"ul\"op}

\vspace{0.3cm}

\centerline{University E\"otv\"os Lor\'and}

\vspace{0.2cm}

\newcommand{\ds}{\displaystyle}

{\large \bf Abstract}

\vspace{0.3cm}

The quark-gluon plasma is written by the non-Abelian gauge theory. The dynamics of the gauge SU(2) are given by the Hamiltonian function, which contains the quadratic part of the field strength tensor $F_{\mu \nu}^a$ expressed in Minkowski space. The homogeneous Yang-Mills equations are solved on lattice $N^d$ considering classical approximation, which exhibits chaotic dynamics. We research the time-dependent entropy-energy relation, which can be shown by the energy spectrum of Kolmogorov-Sinai entropy and the spectra of the statistical complexity.

\section{\bf Introduction}

In order to know the microscopic mechanisms of high-energy physics, non-Abelian gauge theory provides an appropriate theoretical model. It plays an important role in understanding non-equilibrium processes where energy and momentum are in local equilibrium.  Within the framework of perturbative quantum field theory  the  equilibrium and transport processes are studied.

The quantum field theory introduction of quark-gluon plasma was described by the Feynman path integral, which was derived using gauge transformations for non-Abelian gauge fields in a continuous case. The solution of the Yang-Mills equation was determined on a lattice by classical approaching, the equations of motion contain a field strength tensor square.

The time evolution of this system showed to be chaotic\cite{2.,5.,6.}. 
This dynamical quantity is well characterized by the   Kolmogorov-Sinai entropy depending on energy and time resp. the statistical complexity as a function of energy and entropy.

The relation between average energy and Kolmogorov entropy was first introduced in  
\cite{40.} for pure SU(2) Yang-Mills systems.  The finite size extrapolated initial evolution and as a function of the scaled energy was researched in \cite{aftb}. 
The Lyapunov exponent was determined by monodromy matrix and extrapolated ($N\to \infty$),  the scaling properties were studied at a given time range. The spectra of the maximal Lyapunov exponent were calculated depending on the time and energy.
 The Kolmogorov-Sinai entropy was investigated by Pesin form. In this article, we derived the spectrum of this quantity as a function of time and energy.
 
The idea of complexity has been presented several times recently as algorithmic complexity (Kolmogorov) \cite{ank}, the amount of information about the optimal predicts the future according to the expected past (Crutchfield, Young) \cite{cy, av}, finite series complexity (Lempel, Ziv) \cite{lz}.
  
Effective entropy was published by P. Grassberger \cite{gpijt}, taking into account the combination of order and disorder, regularity, and randomness, since most systems do not have the highest Shannon information\cite{33.} (random structure) or the lowest (ordered structures) alone. 
 
The definition of statistical complexity is introduced by R. L\'opez-Ruiz, L. Manchini, X. Calbet (LMC) \cite{39., 37.m, ap} and J. Shiner, M. Davison, P. Landsberg (SDL) \cite{plx}.
 
  The generalized statistical complexity measure (M. Martin, A. Palestino, O. Rosso) \cite{7.} is based on the LMC concept, which describes the finite time series of nonlinear systems together with the associated probability distribution of the dynamic method.  It was extended to Tsallis ($T_q$), Wootters, R\'enyi ($R_q$) \cite{29.} entropy and Kulbach-Shannon, Kulbach-Tsallis, Kulbach-R\'enyi, Jensen divergence. Tsallis suggested generalizing the degree of entropy of the famous Shannon-Boltzmann-Gibbs ($S_{BGS}$) entropy measure\cite{28.}. The new entropy function plays an important role together with its associated generalized thermostats (1998). The Euclidean distance was criticized by Wootters \cite{ww}, who studied this concept in a quantum mechanical context.
   Since the related consideration is an internal statistical measure, this concept can be applied to any probability space.
     Remark that $S_{BGS}$, $T_q$, $R_q$ considered  as special cases of the $(h,\varphi)$ entropies \cite{34.} for the study of asymptotic probability distributions. These quantities were generalized to quantum information theory \cite{34.}. This includes the Neumann entropy \cite{36.} and a quantum version of Tsallis' and Renyi entropies, which have been applied for example to the detection of quantum entanglement \cite{37.b}.

 In addition, we use information theory tools to analyze complex signals, as entropies, distances, and statistical differences play a crucial role in forecasting, estimation, detection, and transmission processes.
This concept has been widely used in the chaotic field \cite{fpp, fa-aip}, symbolic sequences \cite{ac}, pseudo-random bit generator \cite{cghl}, number system \cite{fa-numbs}.

We consider creating a quantitative statistical metric
complexity. It is based on a statistical description of the system
imposing on the physical model.
Suppose the system has $N$ available states $\{x_1, \dots ,x_n\}$
on a given scale and determine the appropriate probabilities
$p_1, \dots ,p_n$ of each state.

The LMC statistical measure of complexity\cite{ap} is described in
two components, i.e. entropy or information stored in the system
and distance from the equilibrium probability distribution,
an imbalance giving its corresponding asymptotic properties
well-behaved measure of complexity.

This quantity is often characterized by a controversial situation
an elaborate dynamics created from relatively simple systems. If
the system itself is already enough contained and it is
consisted of many different parts, you can support a complex dynamic without it
appearance of typical characteristic patterns\cite{entr}. Therefore 
a complex system does not necessarily produce a complex output.

Statistical approaches are easier to implement than to solve
equations of motion, and in many cases offer a solution for treatment
otherwise difficult problems.
 
The structure of the article:
In the second section we introduced the field theory by Feynman path integral and considered the gluonic part of gauge fields. These quantities play an important role in particle and statistical physics. 
In the third section, we discretized these quantities on the lattice by parallel transporter and Wilson action.
In the fourth  we study the maximal Lyapunov spectrum and Kolmogorov-Sinai entropy energy relation.
 In the fifth section, the statistical complexity is introduced on the probability space and we consider the states sequence along the time-evolution of the gauge field, where the state means all links of the lattice at a given time moment.
  
\section{Path integral}

Numerous representations of the field theory are known, Schr\"odinger wave mechanics resp. Heisenberg operator algebra. One of the best-known methods of quantum production with the Feynman path integral. The advantage of this method is that the analogy between statistical physics and particle physics can be easily drawn. It is well applicable to the formulation of the gauge theory and also accurately reproduces its symmetries. This method is briefly described through quantum mechanics.

\subsection{Field theory }

\subsubsection{Feynman  path integral}

Using canonical transformations in classical and quantum mechanics, the action is a general function of the canonical transformation.
 Dirac \cite{8.} applied this procedure in the quantum mechanics to the Hamiltonian function 
  $H$ at time $t$  in the $q'$ state,  respectively.  At the moment $T$ 
for a system where the transient amplitude is:
\begin{eqnarray}
\left< q'_t | q_T \right> \sim \exp\left(\frac{i}{\hbar}\int_T^t L
dt \right). \label{f1}
\end{eqnarray}
On a finite time interval $T-t$, the range $T-t$ is divided into $N$ infinitesimal time intervals, $t_a = t + a\varepsilon$, $N\varepsilon = T-t$. Let $q_a = q_{t_a}$ for all $t_a$. We apply the  $\int dq|q\left>\right<q|=1 = 1$ correlation then  it is following:
$$
\left< q'_t | q_T \right>=\int dq_1 dq_2\dots dq_{N-1} \left<q'_t
|q_1\right>\left<q_1|q_2\right>\dots \left<q_{N-1}|q_T\right>. \label{f2}
$$
The transient amplitude prescribed by the path integral for infinitesimal time interval $\delta t$ introduced by Feynman\cite{rpf}: 
\begin{eqnarray} 
 \left< q'_t | q_{t+\delta t} \right> = \lim_{ {N  \to \infty}
\choose {N \varepsilon  =  konst.}} A^N\int \left(
\prod_{i=1}^{N-1}
 dq_i\right) \left(
\prod_{i=1}^{N-1}
 dp_i\right) \times \nonumber \\ \exp \left(-\frac{i}{\hbar}\int^t_T
 dtL(q,\dot{q})\right),
 \end{eqnarray}
 where $A^N$ is the normalization factor  dividing this coefficient by a factor $A$ for each instant of time.
 This expression is equivalent to the integral of the action function as follows
 \begin{eqnarray}
\equiv  \int {\cal D}q {\cal D}p
 \exp\left(-\frac{i}{\hbar}S(t,T,q)\right). 
\end{eqnarray}
The boundary conditions are the value of orbit at the initial and at the final moment. The above expression gives the probability amplitude of the particle, assuming that it was at $t$ moment in $q'$ state and at time $T$ was in $q$ state. The transient amplitude is expressed as the sum of each of the possible orbits, which begins in $q$ at time $T$ and ends in $q'$  at time $t$, weighted by the exponential expression $(-\frac{i}{\hbar}S)$ for each trajectory.

The expression of the transient amplitude for Hamiltonian systems can be described as 

\begin{eqnarray}
\left<\left. q'_t\right|q_T \right>= \int \dots \int {\cal D} q
{\cal D} p \exp\left(i\int_T^t d\tau \left[p\frac{dq}{d \tau} -
H(p, \left<q\right>)\right]\right),
\end{eqnarray}

where $\left<q\right>$ is the average of $q$ over a given interval.

\subsubsection{Relation between statistical physics and particle physics}\label{rsp}

Statistical mechanics is closely related to the Feynman path integral of quantum mechanics. Creutz showed in 1977 \cite{9.} that the transfer matrix method simplifies the problem of quadratic functions operator diagonalization in Hilbert space.

The Lagrange function of the free nonrelativistic particle, which measure is   $m$  moving in potential $V(x)$ (imaginaries time lattice):
$$
 L(x,\dot{x})=K(\dot{x})+V(x), \quad\quad\quad\quad
K(\dot{x})=\frac{1}{2}m\dot{x}^2.
$$
The action function of any trajectory is following
\begin{eqnarray}
S=\int dt L \left(\dot{x}(t),x(t)\right),
\end{eqnarray}

with which we can specify with the path integral:
\begin{eqnarray}\label{s46}
Z=\int \left[dx(t)\right]\exp({-S}). 
\end{eqnarray}
The integral is for all trajectories $x(t)$. The time component of the lattice is discretized. Investigate the trajectories over the entire $\tau$ time interval,
which is decomposed into discrete time slices of length $a=\tau/N$. The coordinate for the $i$-th time slice is $x_i$. The time derivative $x$ is  approximated  with the difference of the neighbors:
\begin{eqnarray}
S=a\sum_i\left[\frac{1}{2} m \frac{(x_{i+1}-x_i)^2}{a^2} +
V(x_i)\right]. \label{s47}
\end{eqnarray}
Expression (\ref{s46}) is written with $x_i$ coordinates using the $Z$ integral approximation:
\begin{eqnarray}
Z=\int \left(\prod_i dx_i\right) \exp(-S).\label{s48}
\end{eqnarray}
Equation (\ref{s48}) is no different then the shape of partition functions in a statistical physical system.

The procedure that leads from the path integral to the expression of the quantum mechanical Hilbert space in three steps is: First, we define the path integral on a time-like lattice. We construct the transfer matrix in Hilbert space. We finally 
take the logarithm of the transfer matrix, where the linear term expresses the temporal evolution of the system. If the i-th eigenvalue of the transfer matrix is $\lambda_i$, then $Z = \sum\lambda_i^N$. Since the number of time slices goes to infinity, therefore, the expression can be characterized by the maximum self-values $\lambda_0$: 
$$
Z=\lambda_0^N\left[1+O\left(\exp\left[-N\ln\left(\frac{\lambda_0}{\lambda_1}\right)\right]
\right)\right].
$$

\subsection{Gauge fields}\label{gf}

Several introductions of the gauge fields are known. The simplest way is an extension of the Abelian gauge theory describing the electromagnetic field.
The components of the antisymmetric tensor are electromagnetic fields, which are four-dimensional vectors:
$$
F_{\mu \nu}=\partial_{\mu} A_{\nu} -\partial_{\nu} A_{\mu}\quad
\mu,\nu=0,1,2,3.
$$
The space-time indices are denoted by $\mu,\nu$, and the group generators by $\alpha,\beta,\gamma$. Yang and Mills \cite{10.} proposed (1954) to assign the isospin index to $A_{\mu}$ and $F_{\mu\nu}$:
$$
A_{\mu} \rightarrow A_{\mu}^{\alpha} \;\;\;\;\;\;\;\;\;\;F_{\mu
\nu} \rightarrow F_{\mu \nu}^{\alpha}\quad \alpha=1,\dots,N,
$$
 a  further antisymmetric term is added to the expression and the shape of $F_{\mu\nu}$ is:
\begin{eqnarray}
F_{\mu \nu}^{\alpha} =
\partial_{\mu}A_{\nu}^{\alpha}-\partial_{\nu}A_{\mu}^{\alpha}+g_0
C^{\alpha \beta \gamma} A_{\mu}^{\beta} A_{\nu}^{\gamma},
\end{eqnarray}
 where $g_0$ is the bare coupling constant, $C^{\alpha \beta \gamma}$ is the structural constant of the Lie algebra of a $G$ Lie group. Here we use only uniter groups, the fundamental representation of the $G$ group. We parameterize the elements of $G$ with the set of generators $g=\exp(i\omega^{\alpha}\zeta^{\alpha})$, where $\omega^{\zeta}$ is the set of parameters and $\lambda^{\alpha}$ is the set of Hermitian matrices that generalize the group. The structure constants are defined by the following context:
$$
\left[ \zeta^{\alpha},\zeta^{\beta}\right]=iC^{\alpha \beta
\gamma}\zeta^{\gamma}.
$$
The generators are orthonormal:  $\mbox{tr}(\zeta^{\alpha}\zeta^{\beta})=\frac{1}{2}\delta^{\alpha\beta}$. The simplest non-Abelian theory uses the $SU(2)$ group, which is generalized by Pauli matrices
$\zeta^{\alpha}=\frac{1}{2} \tau^{\alpha},\;\;
C^{\alpha \beta \gamma}=\varepsilon^{\alpha \beta \gamma}.$
The Maxwell equations can be derived from the Lagrange density:
$$
{\cal L}=\frac{1}{4}F_{\mu \nu} F_{\mu \nu} + j_{\mu} A_{\mu},
$$
where $j_{\mu}$ is the external source as the electrodynamic fields. The non-Abelian Lagrange density starts in the same way, except for the amount for isospin and $F_{\mu \nu}$ contains an additional member. The classical equation of motion of electrodynamics is the equation $\partial_{\mu}F_{\mu \nu}=j_{\nu}$. In the non-Abelian theory $(D_{\mu}F_{\mu \nu})^{\alpha}=j^{\alpha}_{\nu}$. Here are the covariant derivatives:
\begin{eqnarray} 
(D_{\mu}F_{\mu \nu})^{\alpha} = \partial_{\mu}
F^{\alpha}_{\mu \nu} + g_0C^{\alpha \beta \gamma} A_{\mu}^{\beta}
F^{\gamma}_{\mu \nu}.
\end{eqnarray}
The non-Abelian analog of current conservation following
$$
D_{\mu}j_{\mu}=0.
$$
Second definition of gauge fields uses the local symmetry of the action function. Gauge symmetry of electrodynamics:  $A_{\mu}+\partial_{\mu}\Lambda$, where the gauge function $\Lambda(x)$ is an arbitrary function of the space-time coordinates. In the case of non-Abelian, $A_{\mu}$ is transformed as follows: $A_{\mu}\to g^{-1}A_{\mu}g+\frac{i}{g_0}g^{-1}\partial_{\mu}g$, where $g$ is an element of a suitably chosen group of gauges. In electrodynamics, this transformation is written by a simple phase: $g(x)=\exp({-ig_0\Lambda(x))})$. This is the so-called $U(1)$ gauge theory of electrodynamics. Then, using the transformation formula given above, the transformation of $F_{\mu \nu}$ can be simply written: $F_{\mu \nu} \to g^{-1}F_{\mu \nu} g$. The gauge transformation of the covariance derivative can be given in a similar form.

A third possible introduction to gauge theory is phase theory (Mandelstam (1962)\cite{11.}, Yang (1975)).

In this article we mention the introduction of gauge theories to canonical Hamiltonian formalism following Steven Weinberg (1965) \cite{12.}.

\paragraph{Group}

 In this article, we apply some basic properties of the invariant integral introduced by Haar \cite{14.} in Wilson on compact Lie groups.
 Haar-measure satisfies the following condition:
$$
\int_G f(U)dU= \int_G f(U^{-1})dU.
$$
In the case of $G = SU(2)$ the group elements can be parameterized in the following way:
$$
U=x^0 \mathbf{1} +i\vec{x} \vec{\tau} =\left( \begin{array}{rc}
                                 x^0+ix^3, & x^2+ix^1 \\
                                 -x^2+ix^1, & x^0-ix^3
                              \end{array} \right).
$$
The parameters $x^i$ must be sufficient to satisfy the condition:
$$
\det U=x^2=(x^0)^2+\vec{x}^2=1.
$$
 that specifies the $S^3$ key. In the case of numerical calculation,
we used the quaternion representation $x_0, x_1, x_2, x_3$, because the runtime is faster and the memory requirement is smaller than the matrix representation.

\section{Lattice field theory}\label{lft}

Continuous gauge quantities are introduced on a lattice \cite{1.}. We consider the Wilson action  and the Yang-Mills theory by these discretized quantities.

\subsection{Discrete parallel transporter}

Consider hypercubic lattice of size $a$ and the regularization of the continuous Euclidean lattice. The scalar field $\phi(x)$ is interpreted on the lattice point. Local gauge transformation is following:
$$
\phi(x)\rightarrow \phi'(x)=\Lambda^{-1}(x)\phi(x).
$$
In this case, the nearest non-zero lattice spacing $a$ must be introduced on the hypercube grid.

The elementary parallel transporters are closely connected by the links $b$, which connect the neighboring points. Let $x$ be an arbitrary point on the lattice. Nearest neighbour points can be written in the form $x + a\hat{\mu}$, where $\mu =1,2,3,4$  and $\hat{\mu}$ denotes the $\mu$-th unit vector. The links from $x$ to $x + a\hat{\mu}$ can be denoted by the following ordered pair: $b = (x + a\hat{\mu},x) \equiv (x, \mu)$. The parallel transporter can be described by the link $b$:
\begin{eqnarray}
U(b)\equiv U(x+a\hat{\mu},x)\equiv U_{x \mu} \in G,
\end{eqnarray}
where $G$ is the gauge group. The link thus introduced satisfies the corresponding properties of the parallel transporter. Arbitrary path  $C = b_n \circ b_{n-1} \circ \dots \circ b_1$ corresponds to the parallel transporter $U(b) = U(b_n)\dots U(b_1)\equiv \prod_{b\in C}U(b)$ on lattice, which describes the link variables. These are denoted by $\{U(b)\}$\cite{15.}. Transformation of link variables is following:
$$
U'(y,x)=\Lambda^{-1}(y)U(y,x)\Lambda(x),
$$
where $\Lambda\in SU(N)$ and the size of matrix is $N\times N$.
We define the covariance derivative:
$$
D_\mu
\phi(x)=\frac{1}{a}(U^{-1}(x,\mu)\phi(x+a\hat{\mu})-\phi(x)).
$$
 The term of  derivatives are substituted by  covariate derivatives in the kinetic expression:
$$
\frac{1}{2} \sum_x a^4D_\mu \phi D_\mu \phi =-a^2
\sum_{\left<xy\right>}\phi(x)U(x,y)\phi(y)+4a^2\sum_x \phi(x)^2.
$$
The smallest closed loop on the lattice is called a plaque. A plaque is enclosed by 4 links and it contains the following points: $x, x + a\hat{\mu}, x + a\hat{\mu} + a\hat{\nu}, x+a\hat{\nu}$, denoted by $p = (x; \mu\nu)$. The corresponding parallel transporter can be written in the following form:
\begin{eqnarray}
 U_p\equiv U_{x;\mu \nu}
 \equiv&
 U(x,x+a\hat{\nu})U(x+a\hat{\nu},x+a\hat{\mu}+a\hat{\nu}) \times
\nonumber \\
 &U^{\dagger}(x+a\hat{\mu}+a\hat{\nu},x+a\hat{\mu})U^{\dagger}(x+a\hat{\mu},x),
 \end{eqnarray}
which we call the plaque variables. Wilson's suggestion \cite{16.,17.} is to write the theoretical definition of a simple lattice gauge with the plaque variables:
 $S[U] = \sum_p S_p(U_p)$, that is, the action is summed for all $p$, i.e. $\sum_p=\sum_x \sum_{1\le \mu,\nu \le 4}$ means. The action is written on the elementary plaque (showing only one direction):
\begin{eqnarray}
S_p(U_p)=\beta \left\{1-\frac{1}{N} \mbox{Re} \mbox{tr} U_p \right\}.
\end{eqnarray}

\subsection{Wilson action, lattice Hamiltonian}

Wilson action is gauge invariant quantity because $\mbox{tr}U'_p = \mbox{tr}U_p$ is appropriately chosen for group $SU(N)$, further real and positive. We consider the Yang-Mills action by the Wilson action. We introduced  the vector potential: $A_{\mu}(x)=-igA_{\mu}^b(x)T_b$.   Lie-algebra value vector field was defined on the lattice:
$$
U(x,\mu)\equiv
\exp(-aA_{\mu}(x))=1-aA_{\mu}(x)+\frac{a^2}{2}A_{\mu}(x)^2+\dots
$$
we apply $A_{\nu}(x+a\hat{\mu})=A_{\nu}(x)a\Delta_{\mu}^fA_{\nu}(x)$
where $\Delta_{\mu}^f f(x)=\frac{1}{a}(f(x+a\hat{\mu})-f(x))$.

The Campbell-Baker-Hausdorff expression:

 $\exp({x})\exp({y})=\exp({x+y+\frac{1}{2}[x,y]+ \dots})$  therefore we get:
$$
U_{x;\mu \nu}=\exp\left(-a^2G_{\mu \nu}(x)\right), \quad
\mbox{where}\quad G_{\mu \nu}(x)=F_{\mu \nu}(x)+O(a)
$$
$$
F_{\mu \nu}(x)=\Delta_{\mu}^f A_{\nu}(x)-\Delta_{\nu}^f
A_{\mu}(x)+[A_{\mu}(x),A_{\nu}(x)].
$$
Therefore
$$
1-\frac{1}{N} \mbox{Re}\; \mbox{tr} U_p = 2\mbox{tr} \mathbf{1}
+a^4\mbox{tr}(F_{\mu\nu}(x))^2+O(a^5),
$$
where $Re\; tr$ means the real value of the trace  $U_p$, since $\mbox{tr}G_{\mu \nu}(x)=0$ and $\sum_p \mbox{tr} (F_{\mu \nu}(x))^2=\frac{1}{2}\sum_{x;\mu \nu}\mbox{tr}(F_{\mu \nu}(x))^2$.
 We get the following expression of the Wilson action:
\begin{eqnarray}
S=-\frac{\beta}{4N}\sum_x a^4 \mbox{tr} F_{\mu \nu}(x)F^{\mu
\nu}(x)+O(a^5).
\end{eqnarray}
Because, the leading member coincides with the Yang-Mills action for small $a$ if $\beta=\frac{2N}{g^2}$  and $g$ correspond to the bare coupling constant of the lattice theory. We split the action into time-space components
  \begin{eqnarray}
   S = \frac{2}{g^2}\sum_{p_t}(N-\mbox{tr} U_{p_t}) -\frac{2}{g^2}\sum_{p_s}(N-\mbox{tr} U_{p_s}),
   \end{eqnarray}
where $g$ is the continuous limitation value of the coupling constant, the (-) sign is derived from the Minkovski space-time structure. 
The Taylor series of $U_{p_t}$ is explained in time-dependent term
$$
 U_{p_t} = U(t)U^{\dagger}(t+a_t) = UU^{\dagger} + a_t
U\dot{U}^{\dagger} + \frac{a_t^2}{2}U\ddot{U}^\dagger+\dots,
$$
The expressions appear in the Wilson Action:
$$
 N - \mbox{tr} U_{p_t} = - \frac{a_t^2}{2} \mbox{tr} (U\ddot{U}^\dagger)
 \;\;\;\;\;\;\; O(
   a^3_t) \quad \mbox{correction.}
$$
Since $UU^{\dagger}=1$, trace disappears $N$.
It follows from the first derivative of this term that  $\mbox{tr}(U\dot{U}^{\dagger})=0$ and the second derivative is $\ddot{U}U^{\dagger} +
2\dot{U}\dot{U}^{\dagger} + U \ddot{U}^{\dagger} = 0 $.
Therefore the Hamiltonian lattice action is following:
\begin{eqnarray}
\Delta S_H = \frac{2}{g^2} \left( \frac{a_t^2}{2} \sum_{i} \mbox{tr} \left( \dot{U}_i
\dot{U}_i^{\dagger}\right) - \sum_{ij}\left( N- \mbox{tr} \left(
   U_{ij} \right) \right)  \right). 
\end{eqnarray}
The generalized discretized ansatz can be written:
$$
  S = a_t \sum_t a_s^3 \sum_s L. 
$$
The scaled Hamilton density is able to write in the following form.
 \begin{eqnarray}
   a_t H = \frac{2}{g^2}\left( \frac{a_t^2}{2} \sum_{x,i}
\mbox{tr}\left(\dot{U}_{x,i},\dot{U}_{x,i}^{\dagger}\right) +
\sum_{x,ij} \left(N- \mbox{tr}\left(U_{x,ij}\right)\right) \right),  
\end{eqnarray}
namely
 $$
   H=a_s^3 \sum_s\left( \mbox{tr}\left(\dot{U},\frac{\partial L}{\partial \dot{U}}\right)-L
   \right).
$$
On the lattice, the gauge field can be specified by configuring the link variables. The expected value of the quantity denoted by the  $\{U(b)\}\equiv U$ and $\Theta(\{U(b)\})$ link variables:
\begin{eqnarray}
\left<\Theta\right>=\frac{1}{Z} \int \prod_b dU(b) \Theta
\exp(-S(U)),
\end{eqnarray}
where  $Z=\int \prod_b dU(b)\exp(-S(U))$ and $S(U)$ are Wilson actions. If we introduce $\phi(x)$ the field "material" is given by the corresponding integral:
$$
\left<\Theta\right>=\frac{1}{Z} \int \prod_b dU(b) \prod_x
d\phi(x) \Theta \exp(-S(U,\phi)).
$$
In these expressions, the integration measures $dU(b)$, must be chosen to be gauge invariant. During the gauge transformation it is written:
$$
U'(x,y)=\Lambda^{-1}(x)U(x,y)\Lambda(y)
$$
because the action is invariant: $dU = dU'$, $S(U) = S(U')$.

\subsection{Lattice Yang-Mills theory}\label{lym}

In the following, we use Hamiltonian formulation of the classical lattice $SU(2)$ gauge theory \cite{18.}.  The Hamilton function is considering:
 $$
 H'= \frac{g^2aH}{4} =
\sum_{x,i}\frac{a^2}{4}\mbox{tr}\left(\dot{U}_{x,i}^{\dagger},
\dot{U}_{x,i}\right)+\sum_{x,ij}\left[
 1-\frac{1}{2}\mbox{tr} U_{x,ij}\right], 
 $$
where $U_{x,i}$ is the group element $SU(2)$, this term means the $x+ae_i$ link pointing in the $i$ direction starting at $x = (x_1, x_2, x_3)$ on the lattice. $U_{x,ij}$ denotes the elementary plaque which is expressed by link  $U_{x,ij}=U_{x,i}U_{x+i,j}U_{x+j,i}^{\dagger}U_{x,j}^{\dagger}$  lying in the plane stretched by the elementary vectors $i$ and $j$ starting at $x$.
 We apply the link variables only in the expressions $H$:
 \begin{eqnarray}
H= \sum_{x,i}\left[\frac{1}{2} \langle \dot{U}_{x,i}, \dot{U}_{x,i}
\rangle + \left(1- \frac{1}{4}\langle U_{x,i},V_{x,i} \rangle
\right)\right],
\end{eqnarray}
where the complement link  variable $V_{x,l}(U)$ is following:
$$
V_{x,l}=\frac{1}{4}\sum_{(l,s):{\{(i,j),(k,j),}\choose
{(-i,j),(-k,j)\}}} U_{x+l,s} U_{x+l+s,-l}^{\dagger}
U_{x+l,-l}^{\dagger},\quad \mbox{where}
$$  
$i,j,k$ are the unit vectors of the three-dimensional lattice.

In the gauge field section (\ref{gf}) we introduced the quaterinon representation, which is defined in the following way on a lattice:
 \begin{eqnarray}
 U  =  u_0 \mathbf{1}+i\vec{\tau} \vec{u} \quad \quad
 U   =   \left( \begin{array}{c}  u_0+iu_3, iu_1+u_2 \\
                                  iu_1-u_2, u_0-iu_3  \end{array}
                                                   \right).
                                                   \nonumber
 \end{eqnarray}
The equations of motion are derived from the Hamiltonian function:
\begin{eqnarray}
\dot{U} & = & P, \nonumber \\
\dot{P} & = & V-\left<U,V\right>U-\left<P,P\right>U,
\end{eqnarray}
where  $\left<P,P\right>=\frac{1}{2} \sum_jP_jP^j$.

The lattice equation of motion \cite{19.} follows:
 \begin{eqnarray}
 U_{t+1}-U_{t-1} & = & 2h(P_t^{\star}-\varepsilon U_t^{\star}),
\\
 P_{t+1}-P_{t-1} & = & 2h( V(U_t^{\star})-\mu U_t^{\star}+\varepsilon P_t^{\star}
 ), \quad \mbox{where}
\nonumber
 \end{eqnarray}
 $$
 \varepsilon = \frac{ \langle U_t^{\star},P_t^{\star} \rangle}{\langle
U_t^{\star},U_t^{\star} \rangle},\quad \quad \quad \quad \mu =
\frac{\langle V(U_t^{\star}),U_t^{\star} \rangle + \langle
P_t^{\star}, P_t^{\star} \rangle}{ \langle
 U_t^{\star},U_t^{\star} \rangle}, \quad \mbox{and}
$$
$$
U_t^{\star}=aU_{t+1}+bU_t+cU_{t-1}.
$$
The quantities $\varepsilon, \mu$ denote the Lagrange multipliers.
The energy of the Hamiltonian system was constant and Gaussian law is satisfied \cite{3.} during the movement. A periodic boundary condition was used to solve the system of equations. The color charge was defined following:
$$
\Gamma_i=\sum_{l_+}P_lU_l^\dagger-\sum_{l_-}U_l^\dagger P_l, \quad
i=1,\dots N
$$
The measure of change is written by this term:
$$
\dot{\Gamma}_i =
\sum_{l_+}(VU^\dagger-\left<V,U\right>\mathbf{1}),
$$
where $P_1 = QU_1$ and $P_1=QU_1$ and $P_n=U^{\dagger}_{n-1}P_{n-1}U_n$, $1< n <N$. The condition of neutrality formulated as
$$
Q-F^\dagger QF=0, \\
\mbox{tr} Q=0,
$$
from which it follows
$$
Q=\frac{q}{2}(F^\dagger-F), \quad \mbox{where} \quad
F=\prod_{i=1}^{N-1}U_i \quad \mbox{oriented product}
$$
the initial color charge is $Q$ and the final state is $-F^\dagger QF$.

\section{Nonlinearity}

In this section, we numerically determined the Lyapunov spectrum on the three-dimensional lattice of the SU(2) Yang-Mills field. The spectra of Kolmogorov-Sinai entropy are studied by the eigenvalues of the monodromy matrix from the classical chaotic dynamics to  extrapolate on a lattice with a large size limit.

\paragraph{Monodromy matrix}

We consider a periodic orbit of the energy $E$, with initial phase space coordinates $(p=p_0, x=x_0)$ and final coordinates ($p=p_0$, $x=x_0$). 
We study the behavior of the neighborhood path of the periodic orbits, 
  how these trajectories develop in the case of small transverse perturbation.

This means the same situation when considering the deviation of flow on the Poincare surface of the section transverse to the path.
Then the relation between the initial $\{\delta y_{0i},\delta p_{0i}\}$  and final state $\{\delta y_i,\delta p_i\}$ deviation is following:
$$
\delta y_i=\sum_{j=1}^{d-1} \left(\frac{\partial y_i}{\partial y_{0i}}\right)\delta y_{0j} + \left(\frac{\partial y_i}{\partial p_{0i}}\right)\delta p_{0j}= \sum_{j=1}^{d-1} A_{ij}\delta_{0j}+C_{ij}\delta p_{0j}
$$
and
$$
\delta p_i=\sum_{j=1}^{d-1} \left(\frac{\partial p_i}{\partial y_{0i}}\right)\delta y_{0j} + \left(\frac{\partial p_i}{\partial p_{0i}}\right)\delta p_{0j}= \sum_{j=1}^{d-1} C_{ij}\delta_{0j}+D_{ij}\delta p_{0j}
$$
It is written by  matrix form: 
\begin{eqnarray}
\left(\frac{\delta \overline{y}}{\delta \overline{p}}\right) = \left(\begin{array}{cc}
\overline{A},& \overline{B} \\
\overline{C},& \overline{D}
\end{array} \right) \left(\begin{array}{c}
\delta \overline{y}_0\\ 
\delta \overline{p}_0
\end{array} \right)
= \overline{M}  \left( \begin{array}{c}
\delta \overline{y}_0\\ 
\delta \overline{p}_0
\end{array} \right),
\end{eqnarray}
where $\delta \overline{y}$ and $\delta \overline{p}$ are $1 \times (d-1)$ dimensional column matrices, and $\overline{A}, \overline{B}, \overline{C}$, $\overline{D}$ are $(d-1)\times (d-1)$ dimensional square matrices where  $A_{ij},B_{ij},C_{ij}$  $D_{ij}$  matrix elements. This $(2d-2)\times (2d-2)$ dimensional square matrix $\overline{M}$ means the monodromy matrix  according to the equation motion \cite{21.}. 

The shape of the monodromy matrix by the lattice equations of motion \cite{aftb} is following 
\begin{eqnarray}
M=
\left( \begin{array}{cc}
\frac{ \partial \dot{U}}{ \partial U}  & \frac{ \partial \dot{U}}{ \partial P} \\
  \frac{ \partial \dot{P}}{ \partial U} & \frac{ \partial \dot{P}}{\partial P}
\end{array}
\right).
\end{eqnarray}
We write down each partial derivative by the equation of motion:
$$
\begin{array}{c}
\frac{  \partial \dot{U}^a}{ \partial
U^b}=0,\nonumber
\end{array}\quad \quad \quad
\begin{array}{c}
\frac{ \partial \dot{U}^a}{ \partial P^b}= \delta^{ab},
\nonumber
\end{array}
$$
$$
\begin{array}{c}
\frac{\partial \dot{P}^a}{  \partial U^b} =\frac{
\partial V^a}{ \partial U^b}-\left(\sum_{c=1}^N U_c\frac{ \partial
V^c}{ \partial
U^b}\right)U^a-V^bU^a-\sum_{c=1}^N\left(U_cV^c+P_cP^c\right)\delta^{ab},
\end{array}
$$
$$
\begin{array}{c}
\frac{\partial \dot{P}^a}{ \partial P^b} = -2P^bU^a, \quad \mbox{where}
\nonumber
\end{array}
$$
$$
\frac{ \partial V_k^{\alpha_q}}{ \partial U^{\beta_q}} =
\sum_{l=1}^{\cal N} \frac{ \partial
V_k^{\alpha_q}(U_1,\dots,U_{\cal N})}{ \partial U_l^{\beta_q}}, \quad \mbox{where ${\cal N}=12, \quad \alpha_q, \beta_q=0,1,2,3$.}
$$
The shape of the characteristic equation is then:
\begin{eqnarray}
\det \left[ \left( \begin{array}{cc}
\mathbf{0}  & \mathbf{1} \\
  \frac{ \partial \dot{P}}{ \partial U} & \frac{\partial \dot{P}}{ \partial P}
\end{array}
\right)- \Lambda_i \mathbf{1} \right] =0.
\end{eqnarray}
We showed the stability of the trajectories along the trajectory in the vicinity of any point on the $(U,P)$ phase space. The time evolution of a small $(\delta U, \delta P)$ perturbation is determined by the monodromy matrix. Among the eigenvalues of the stability matrix, real and positive quantities indicate an exponential departure of adjacent trajectories, i.e., motion is unstable. At the long-term limit, the Lyapunov exponents are obtained from the eigenvalues.

\subsection{Spectrum of the maximal Lyapunov exponent}

We investigated the ergodization of the SU(2) lattice gauge theory due to classical chaotic dynamics \cite{aftb}. 
We get a good approximation to the real maximum Lyapunov spectrum by monodromy matrix of time-evolving field configurations. The lattice size was chosen to be $N = 2,3,4,5,6,7$. 
The initial configurations are randomized we choose according to the Haar measure and the total energy constraint.

 The Lyapunov exponent $L_i$ is introduced with eigenvalues $\Lambda_i$ of monodromy matrix:
\begin{eqnarray}
L_i = \lim_{T \rightarrow \infty} \frac{\int_0^T 
\Lambda_i(t) dt}{T} \quad i=1,\dots,f,
\end{eqnarray}
where $\Lambda_i(t)$ is the solution of the characteristic equation: 
\begin{eqnarray}
\det[\Lambda_i(t) \mathbf{1}-M(t)] = 0, 
\end{eqnarray}
in which $M$ is the linear stability matrix, $f$ is the number of degrees of freedom. Conservative dynamical systems satisfy the Liouville theorem:
$
\sum_{i=0}^f L_i=0.
$
In numerical calculations, we use the definition of the discrete Lyapunov spectrum
\begin{eqnarray}
L'_i = \left<\Lambda_i\right>^{(n)}=\frac{1}{n}\sum_{j=1}^n
\Lambda_i(t_{j-1}), \quad i=1,\dots,f,
\end{eqnarray}
where $t_j$ is the time series during the trajectory evolution of the gauge field configuration. 

 The quantities $L'_i$  are extrapolated to a long-term $(N \to \infty)$ limit with fixed time steps. We assumed it converges to the $L_i$ Lyapunov exponent in noncompact configuration space.

 The eigenvalues of the monodromy matrix were determined along the time-evolution of a single gauge trajectory which allows us to know the behavior of the Lyapunov spectrum as a function of time.  
 
In the numerical simulation the total number of degrees of freedom 
$f = 4 \times 3 \times N^3 = 12 \times N^3$, where the group element $SU(2)$ is represented by 4 real quaternions (thus the phase space has a dimension of $2f = 24N^3)$. Due to the conditions of survival (unity, orthogonality), the number of physically relevant degrees of independent freedom decreases \cite{22.}.

The spectrum of the $2f \times 2f$ stability matrix although rare is large enough to determine the eigenvalue with sufficient accuracy.  Since it requires $O(f^2)$ memory to calculate eigenvalues, $N = 7$ ($2f=24N^3=8232$ dimensional phase space) was the maximum size of the system, which could be examined by the capacity of the computer, which is due to the fact that the Hamiltonian system is conservative (energy is time-independent).

In the literature, it has been shown that in the semiclassical limit the real-time
 Hamiltonian dynamics of SU(2) gauge theory exhibits deterministic chaos on a spatial lattice  \cite{2.}. The largest Lyapunov exponent of the gauge field was calculated as a function of energy density.  Numerical integration of the equations of motion has been applied  considering the conservation of energy and Gaussian law. 
The exponential divergence of two trajectories was studied on the lattice gauge field configuration. The gauge-invariant metric is proportional to the absolute local difference in the magnetic energy of two different gauge fields. The nearest neighboring configurations were chosen randomly and along the time-evolution, the distance between the two trajectories increased exponentially until it is saturated. This process is known as the rescaling method.

 In this paper, we determine the maximum value of the Lyapunov exponent along with the real-time evolution of a single long trajectory using the monodromy matrix.
Our goal is to calculate the spectrum of maximal Lyapunov exponent depending on the energy resp. time and we consider the scaling behavior of this system.

Therefore the first step we extrapolated the real maximal Lyapunov exponent ($N\to \infty$) to the thermodynamical limit from the dataset, which is taken for $N=2,3,4,5,6,7$ at the different energies $g^2aE\in [0.0,0.7]$ range considering the finite-size scaling.  

\begin{figure}
\centering
\includegraphics[width=11.5cm]{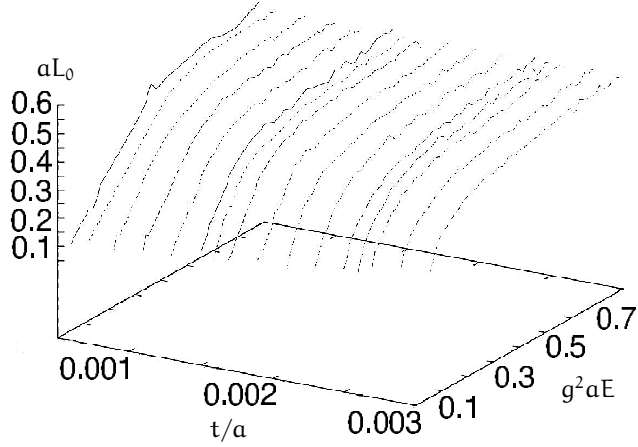}
\scalebox{0.8}{ { \put(-337,230){\makebox(0,0)[t]{$aL_0
$}}}}
\scalebox{0.8}{
\put(-260,60){\makebox(0,0)[t]{$t/a$}}}
\scalebox{0.8}{
\put(-95,80){\makebox(0,0)[t]{$g^2aE$}}}
\caption{Maximal Lyapunov spectrum $aL_0$ as a function of the scaling time $t/a$ and scaling energy $g^2aE$.}\label{fig-1}
\end{figure}

Figure (\ref{fig-1}) shows the real maximal Lyapunov exponent's $aL_{0}$ dependence on scaling time $t/a$ and scaling energy $g^2aE$, where $a$ is a lattice size and $g$ means the strong coupling constant (Section \ref{lft}.).

The scaling of the maximal Lyapunov exponent as a function of scaling energy has  been studied \cite{aftb}. In the past, the research on the scaling behavior of maximal Lyapunov exponent has been debated whether it is linear or not in the long-time limit\cite{aftb}. According to some research results, this would be  $L_0 \sim E^{\frac{1}{4}}$  relation. 
It has been shown that linear scaling at low energy is acceptable using the rescaling method in the long-term boundary case. 

In the Figure (\ref{fig-1}) the scaling of the maximal Lyapunov exponent  at short time range  $t/a=0.0005$ satisfies the linear $L_0\sim E$ relation before the curve saturates. In the long-time limit   at $t/a=0.003$ the scaling becomes logarithmic rather than $L_0 \sim E^{\frac{1}{4}}$ relation \cite{aftb}. It can be considered that too long a trajectory and the compactness of the configuration space create the calculated eigenvalues, which is the Hamiltonian lattice field theory artifact. In the following, we imply linear scaling.

The extrapolation of the maximum Lyapunov exponent values was plotted on the Figure (\ref{fig-1}) i.e. the thermodynamic limit $N\to \infty$ at different energies. The finite-size scaling of this quantity to be almost linear:
$$
L_0 \sim \frac{1}{\sqrt{f}} \sim N^{-\frac{3}{2}}.
$$
 This corresponds to sampling ergodic states \cite{23.}.

 \subsection{Spectrum of Kolmogorov-Sinai Entropy}

The relation between average energy and Kolmogorov-Sinai entropy was first published in  \cite{2.} for the simple $SU(2)$ Yang-Mills system.

We define the Kolmogorov-Sinai entropy by the term Pesin:
\begin{eqnarray}
h^{KS}=\sum_i L_i\Theta(L_i),
\end{eqnarray}
where the value of the function $\Theta(x)$ equals 1 if the argument is positive and 0  otherwise. The dimension of the quantity $h^{KS}$ is a rate (1/time). Therefore, the  entropy  can be given on an $N^3$ lattice by normalizing quantity:
\begin{eqnarray}
 S =\frac{h^{KS}}{\mbox{Re}(L_0)N^3}.
\end{eqnarray}
The state equation can be derived from the simulations of the dynamics. The finite-size scaling is  extrapolated to infinity $(\frac{1}{N} \to 0)$ on the lattice.
We consider the Kolmogorov-Sinai entropy as a function of time and energy. This leads to the state equation, which is the relation of entropy-energy $S(E)$ in the thermodynamic limit of infinite volume.

\begin{figure}
\centering
\includegraphics[width=11.5cm]{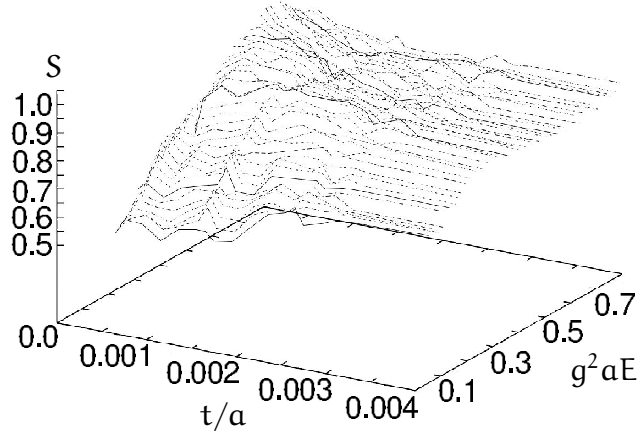}
\scalebox{1.0}{ { \put(-270,180){\makebox(0,0)[t]{$S
$}}}}
\scalebox{1.0}{
\put(-210,40){\makebox(0,9)[t]{$t/a$}}}
\scalebox{1.0}{
\put(-70,67){\makebox(0,0)[t]{$g^2aE$}}}
\caption{Entropy spectrum $S$ depends on the scaling time $t/a$ and scaling energy $g^2aE$.}\label{fig-2}
\end{figure}

 The normalized Kolmogorov-Sinai entropy is derived from the extrapolated $L_i$ data, which depends only slightly on the initial values and scaling linearly according to the energy. 
$$
\left<S\right> \sim b\lg (g^2Ea)+ c,
$$
where $b,c\in \mathbb{R}$.
This is an appropriate estimation of the inverse temperature:
$$
\frac{1}{T}=\frac{\partial \left<S\right>}{\partial E}\sim \frac{0.5}{E}
$$
Thus the equipartition, i.e. the energy per degree of freedom:
$$
E = \frac{1}{2}k T.
$$
In the Figure (\ref{fig-2}) the entropy spectrum $S$ depending on the scaling time$t/a$ and scaling energy $g^2aE$ is plotted on the ranges $t/a\in [0,0.004]$, $g^2aE \in [0,0.7]$. The closest relation of the entropy $S$ as a function of scaling energy $g^2aE$  is the ideal gas $S \sim \lg E$ within the interval of scaling time $t/a$ [0.001,0.004]. In the short range of the scaling time $t/a$ $[0,0.001]$ the lattice artifact appears.

Since the Kolmogorov-Sinai entropy was determined from the Lyapunov exponents with the Pesin form, the lattice artifact experienced in the numerical calculation of the Lyapunov exponents manifests in the Kolmogorov-Sinai entropy spectrum. 

It has been shown that the entropy of the SU(2) lattice gauge field has a first-order phase transition \cite{24.}. The entropy as a function of energy was expressed by the action on the microcanonical ensemble (section \ref{rsp}).

In our case lattice SU(2) system $S(E)$ curve would show a first-order two-phase structure containing a break somewhere or crossover (two-phase structure) at the range of time [0.001,0.003] on the interval of the energy [0.1,0.6]. To decide this, we need to filter out lattice artifacts and reduce entropy fluctuations to give a clear answer. The numerical error can be derived by maximal Lyapunov exponents determination, resp. calculation of the eigenvalue of rare matrices.

\section{Spectrum of Statistical Complexity}

\subsection{Statistical Complexity}\label{stat-compl}

The family of statistical complexity measures $C$ is introduced by the functional product form $C=H\cdot Q$ for difference disorder $H$ and disequilibrium $Q$ measures on the probabilistic space\cite{7.}.

The information measure $\mathcal{L}$ is able to be described by a given
probability distribution $P = \{p_j, j = 1,\dots,n\}$, and this quantity
corresponds to the measure of uncertainty of a physical system. The amount
of disorder $H$ is defined:
\begin{eqnarray}\label{h}
H[P] = \mathcal{L}[P]/\mathcal{L}_{max}, 
\end{eqnarray}
where $\mathcal{L}_{max} = \mathcal{L}[P_e]$ and $P_e = \{1/n, \dots, 1/n\}$ is the uniform distribution which maximizes the information measure $(0 \ge H \ge 1)$.
 
To take into account the idea of statistical complexity, a disequilibrium $Q$ needs to be identified.

The measure of this quantity is examined at some distance $D$ to the equal probability distribution $P_e$.
\begin{eqnarray}\label{d}
Q[P] = Q_0D[P, P_e], 
\end{eqnarray}
where $Q_0$ is a normalization factor $(0 \le Q \le 1)$.
This concept describes the structure of systems as larger than zero if there are possibly more steady states among the possible situations.

Therefore, we take the following functional form for the statistical
complexity measure:
\begin{eqnarray}\label{c}
C[P] = H[P] \cdot Q[P]. 
\end{eqnarray}
This quantity $C[P]$ characterizes the amount of information stored and its disequilibrium in this system altogether \cite{37.m}. 
The definition of this concept can be divided into three categories:
(c1) this quantity  increases monotonically as the function of entropy;
 
(c2) it is a convex function that contains the maximum value of $C_ {max}$ for the probability distribution $P_e$ and the minimum value of $C_ {min}$ that occurs at the extreme values of entropy, i.e. $H=0$ or $H=1$;

(c3) the third type decreases monotonically with increasing entropy \cite{37.m}. 

The two extreme situations can be understood as follows:

 (i) Each set of sequences has the same probability distribution. All of them accept the information stored in an equal measure similar to the ideal gas\cite{38.}.

The probability distribution is the same for all series. All of them accept the information stored in the equivalent measure as the ideal gas \cite{38.}.

 (ii) If we research a system with certain symmetry properties and distance, then this object is able to write by minimum information as a mineral or symmetrical in quantum mechanics or the system is completely disordered.

The statistical complexity is characterized by the scale because it was introduced in a finite system. At each scale of measurement, a new set of
available simulated series occurs with its appropriate probability
distribution $P$; so the complexity is changing.

In statistical mechanics, isolated systems often occur that have arbitrary initial conditions and a discrete equal probability distribution \cite{39.}. It was concluded that in the case of time-evolving isolated systems and their statistical complexity, the measurements should not take arbitrary values in the $C_{LMC}$ as a function of $H$. These constrain the bounds of complexity to certain limits of minimum and maximum value.

We use the Shannon entropy measure and
Euclidean distance on the probability space as the statistical complexity was investigated by Lopez-Ruiz, Manchini, and Calbet (LMC)\cite{37.m}.

\paragraph{Information measure}
We consider the Shannon logarithmic information
measure on the $P \equiv \{p_1, \dots, p_n\}$ discrete probability distribution in this article as follows:
\begin{eqnarray}
\mathcal{L}[P]=-\sum_{j=1}^np_j\log(p_j)
\end{eqnarray}
The maximal value $\mathcal{L}_{max}$ is calculated by the uniform probability
$P_e = \left\{\frac{1}{n}, \dots, \frac{1}{n}\right\}$ fulfilling this criterion $\sum_{j=1}^n p_j=1$ so, $\mathcal{L}_{max}=\ln n$.
If $\mathcal{L}[P] = 0$, it means that the possible outcomes
$j$ whose probabilities are given by $p_j$ will currently take place.
The knowledge of the advantaged process is corresponded by the probability
distribution, in this case, is maximal.
Anyway this quantity turns into largest for a uniform distribution, when $\mathcal{L}[P] = S_{max}$. These two extreme criteria correspond to the (i) perfect order and (ii)maximum randomness as trivial ones. 

\paragraph{Disequilibrium}
Evidently, the Euclidean statistical distance is taken to
give the quantity $D$, i.e., the quadratic distance between the probability
distributions of each state to the equiprobability. If $D$ means the
Euclidean norm in $\mathbb{R}^n$, we find
\begin{eqnarray}
D_E[P,P_e]=\parallel P-P_e\parallel=\sum_{i=1}^n(p_i-p_e)^2,
\end{eqnarray}
where $p_e = 1/n$.
The maximum disequilibrium is gained for overwhelming simulation
sequences with $p_i \sim 1$ and $D \to 1$ for increasing $n$, as long as this
quantity disappears $D \sim 0$ for $p_i \sim 1/n$ for all $i$. In other probability distribution, the value of the disequilibrium $D$ will vary
between these two extreme rates. Then, the expression of the normalization
factor of the Euclidean statistical distance fulfills $Q_0 = \frac{n}{n-1}$.

\subsection{Complexity of the lattice Yang-Mills equation}

In the section (\ref{stat-compl}) we introduced the statistical complexity which is based on the probability distribution providing a statistical estimation of the series of dynamical systems. There are $n$ finite different elements on the sequence $\{x_1, x_2, \dots , x_n\}$ corresponding to the set of discrete probability distribution $P \equiv \{p_1, p_2,\dots , p_n\}$, where $p_i := P(x_i)$, $(\sum_{i=1}^n p_i = 1)$, and $p_i > 0$ for all $i$.

We study the real-time evolution of the gauge field by the Yang-Mills equation on the lattice. Random initial values are chosen which fulfill the constraint (unitarity, orthogonality, and energy). The length of trajectory is taken as $n=10000$, the subsequent along the orbit is $m=2$. The lattice size was chosen $N=2,3,4,5,6,7$. 

The state of the gauge field at time $t$ contains all $U_{x,i}$ links on a lattice of size $a$. The number of links is $dim*N^3$. The lattice gauge field configuration characterizes the state at a given time instant by the links altogether.

The value of entropy (\ref{h}), disequilibrium (\ref{d}), and
the statistical complexity (\ref{c}) can be calculated by the simulation unambiguously.
Since the probability distribution of element 
is discontinuous in three-dimensional lattice gauge space, some complexity
and disequilibrium values do not appear for certain entropy
quantities.

\begin{figure}
\begin{center}
\includegraphics[width=0.8\textwidth]{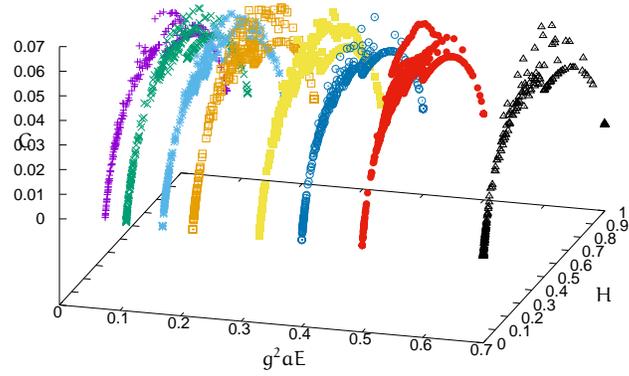}
\scalebox{0.65}{ { \put(-425,170){\makebox(0,0)[t]{$C
$}}}}
\scalebox{0.65}{
\put(-280,45){\makebox(0,0)[t]{$g^2aE$}}}
\scalebox{0.65}{
\put(-100,80){\makebox(0,0)[t]{$H$}}}
\caption{left:Complexity spectra $C$ as a function of the  $H$ and $0<g^2aE<1$ (0.075,0.11,0.17,0.22,0.33,0.4,0.5,0.7) and the lattice size  $N=7$,$m=2$.}\label{fig-3l}
\end{center}
\end{figure}

In the Figure (\ref{fig-3l}) the complexity $C$ as a function of scaling energy $g^2aE$ and entropy $H$  is presented and the lattice size is $N=7$. The spectrum of complexity $C$ was calculated for 8 different energy values, $g^2aE = 0.075,0.11,$ $0.17,0.22,$ $0.33,0.4,0.5,0.7$. The spectrum of complexity $C$  is finite and limited but not necessarily a unique function of entropy $H$ and there exists a convex boundary and larger internal structure between the minimal value $C_{min}$ and the maximal value $C_{max}$ for different energy range [0.075,08].  The minimal and maximal boundary is increasing as the energy is growing. 

The eight different spectra of the statistical complexity $C$ as a function of entropy $H$ and energy $g^2aE$  are determined with the same dynamics, i.e. their internal structure leads to a similar probability distribution along time-evolution.

\begin{figure}
\center{\includegraphics[width=0.8\textwidth]{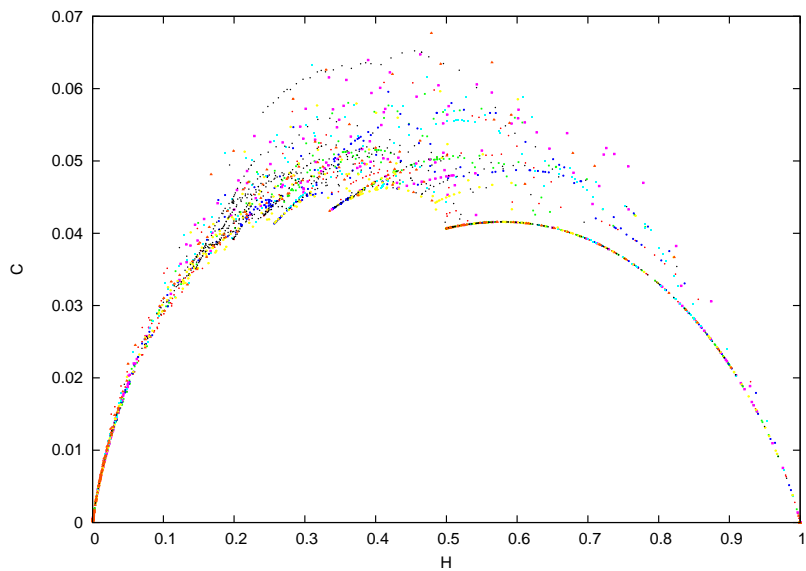}}
\caption{$C$ depends on $H$ on the lattice size $N=7$,$m=2$ for eight different energy rates.}\label{fig-3r}
\end{figure}

The inner structure can be seen better in Figure (\ref{fig-3r}), where the complexity $C$ dependence on the entropy $H$ is shown for eight different energy rates.  The value of the complexity $C$ becomes to zero at $H\sim 0$  and $H \sim 1$ and the curve is convex on the interval $H \in [0,1]$.  This behavior of complexity belongs to a class (c2).  In the immediate neighbour of the $C_{max}\sim 0.07$ values for entropy $H \sim 0.5$, i.e. near to the equilibrium distribution $P_e$, the values of complexity are more strongly scattered than in the case of $H \sim 0$ or $H \sim 1$.

As we have seen in these Figures  their lower boundary $C_{min}$ shows slightly scattered curves with decreasing entropy values, where the maximum value of each curve increases in proportion to the energy in the range entropy $H$ [0.5,1.0]. The upper bound values of the complexity $C$ are widely scattered in the neighbor of equilibrium distribution $P_e$. On the interval of the entropy [0,0.5], the figure does not show any internal structure, where $C_{max}$ and $C_{min}$ belonging to the dynamics of the Yang-Mills system assume almost the same value.

\begin{figure}
\center{\includegraphics[width=0.9\textwidth]{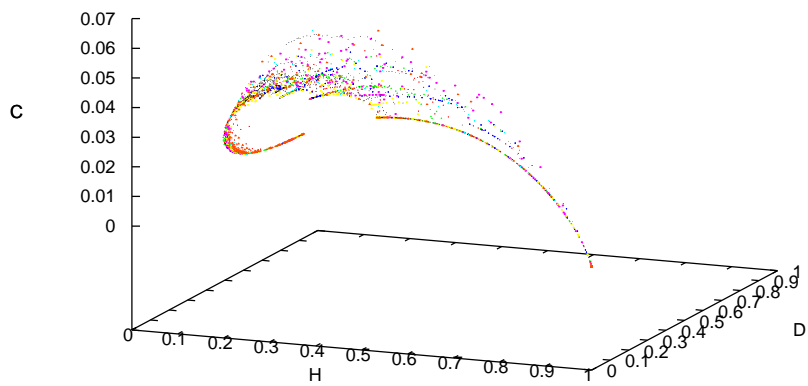}}
\caption{left:Complexity $C$  as a function of the $H$ and $D$, $0<g^2aE<1$ on the lattice size $N=7$,$m=2$ for eight different energy rates.}\label{fig-4l}
\end{figure}

In the Figure (\ref{fig-4l}) the complexity $C$ as a function entropy $H$ and disequlibrium $D$  is plotted. 
 Because the number of points on a long trajectory is finite, $C$ is a function $H$ shows scaling behavior, i.e., the bigger complexity appears at less entropy with a larger discrete probability distribution. Due to the symmetry SU(2) of the non-Abelian gauge field and the constraint of the total energy and Gaussian law, the system does not reach all states of phase space. In constract to the  ideal gas \cite{38.}, where  all state of phases space was available, the internal structure  evenly filled in the range between the $C_{max}$ and $C_{min}$ boundary.  
 
The statistical complexity of the non-Abelian gauge theory was studied  for a long time evolution along the trajectory $n$. It is showed  internal structure in the immediate vicinity of the equilibrium distribution $P_e$, the further research  allows us to narrow the energy range to be examined for the immediate vicinity of the entropy, because the $S(E)$ curve would present a first-order two-phase structure i.e.  having a break somewhere within a certain energy range, that 
the lattice artifact could be filtered out.

\section{Summary} 
 
In this article, we considered the Hamiltonian function on lattice gauge theory in especially the maximal real Lyapunov spectrum of the non-Abelian gauge theory.
The spectra of Kolmogorov-Sinai entropy were studied as a function of energy and lattice size approaching the thermodynamical limit for SU(2) lattice gauge theory.
 Long time evolution of the equation of motion of gauge fields was characterized by statistical complexity in a probability space. The inner structure of this quantity as a function of entropy allows a more accurate determination of the phase transition in non-Abelian SU(2) lattice space theory using a monodromy matrix with appropriate parameter range on growing lattice size by eliminating the effect of the lattice artifact.
 
\pagebreak

\end{document}